\documentclass[preprint,showpacs,showkeys,preprintnumbers,amsmath,amssymb]
{revtex4}

\usepackage{graphicx}
\usepackage{dcolumn}
\usepackage{bm}

\begin{document}

\title{Classical and quantum coupled oscillators: symplectic structure}  

\author{A. R. Bosco de Magalh\~aes}
\email{arbm@fisica.ufmg.br}
\affiliation{Departamento de F\'{\i}sica, ICEX, Universidade Federal
de Minas Gerais, C.P. 702, 30161-970 Belo Horizonte, MG, Brazil}

\author{C. H. d'\'Avila Fonseca}
\email{cazeitor@fisica.ufmg.br}
\affiliation{Departamento de F\'{\i}sica, ICEX, Universidade Federal de
Minas Gerais, C.P. 702, 30161-970 Belo Horizonte, MG, Brazil}

\author{M.C. Nemes}
\email{carolina@fisica.ufmg.br}
\affiliation{Departamento de F\'{\i}sica, ICEX, Universidade Federal
de Minas Gerais, C.P. 702, 30161-970 Belo Horizonte, MG, Brazil}

\date{\today}

\begin{abstract}
We consider a set of $N$ linearly coupled harmonic oscillators and show that
the diagonalization of this problem can be put in geometrical terms. The
matrix techniques developed here allowed for solutions in both the classical
and quantum regimes.
\end{abstract}



\maketitle

\section{Introduction}

Quantum mechanics is a theory constructed to deal with isolated systems.
However, it is impossible to completely isolate a system from its
surroundings, and usually it relax towards thermal equilibrium with the
environment, a phenomenon called dissipation. Otherwise, the environment
acts over the system due to its thermal fluctuations, leading to the
destruction of pure quantum states in a much shorter time scale; this
phenomenon is called decoherence \cite{art1,art2,art3}.

Nowadays, the main theoretical proposal for describing dissipation and
decoherence phenomena is to consider the system of interest plus the
environment as a closed system and to determine the effective dynamics of
the system of interest. This may be done exactly by projection technics 
\cite{art4},
and the resulting equation is very useful for an abstract knowledge about
the structure of the problem. However, since the exact equation that governs
the dynamics of a quantum system is non Markovian and non linear, to obtain
its solution is possible only in some very particular cases.

A model which has been very explored in the literature is the linearly
coupled oscillators model proposed by A. Caldeira and A. J. Legget 
\cite{art5,art6}.
In this model, the system of interest is a harmonic oscillator and the
environment is modeled as huge set of oscillators in thermal equilibrium in
a given temperature; the system-environment interaction is given by a linear
coupling in the position of the oscillator of interest. The Caldeira-Legget
model has been very successful in various areas of physics, particularly in
quantum optics. As is well known, despite its quantitative success in the
analysis of several experiments \cite{art7,art8}, 
a sound derivation and understanding
of the hypothesis involved in its derivation remain to this day an open
question.

The objective of the present contribution is to understand in detail the
classical and quantum symplectic structures of a system composed by an
oscillator linearly coupled to $N-1$ others without tracing out the $N-1$
states which represent the environment. In particular we show that the exact
problem possesses a simple physically transparent geometrical structure. In
Section 2, we treat the classical oscillators. The $N=2$ case is analyzed
first; this problem, although very simple, is structurally rich, and clearly
exhibits the geometrical nature of the solution: we analyze sections in the
phase space which guide us in a sequence of variables transformations (a
squeezing, a rotation and another squeezing) that give us the system's
normal modes. These transformations encountered for $N=2$ are then used to
find analogous transformations for the generic $N$ case. In Section 3, we
work with the quantum oscillators. The transformations encountered for the
classical case are used there in order to find and analyze the normal modes
of the system. Throughout this work we use matricial technics which have
been very useful to express the algebra involved here in a clear and as
short as possible form. In Section 4, we comment on our results and their
relations with modern problems.

\section{Classical Systems}

\bigskip

\subsection{Two Coupled Oscillators}

\begin{figure}[!ht]
\includegraphics[width=16cm]{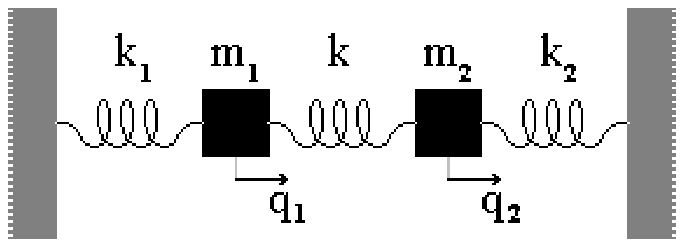}
\vspace{-10cm}
\caption{Two linearly coupled oscillators.}
\label{OA}
\end{figure}

Let us consider the two coupled oscillators' Hamiltonian 
\begin{equation}
H=\frac{1}{2}\omega _{1}\left( p_{1}^{2}+q_{1}^{2}\right) +\frac{1}{2}\omega
_{2}\left( p_{2}^{2}+q_{2}^{2}\right) +g\ q_{1}q_{2},  \label{hamitoniano1}
\end{equation}
where $q_{i}$ and $p_{i}$ are canonical conjugate variables, $g$ and $\omega
_{i}$ are real numbers, $g<0$ and $\omega _{i}>0$. This may be concerned
with the system represented in Fig. (\ref{OA}) for example, where we have
two oscillators with masses $m_{i}$ and elastic constants $k_{i}$ coupled
(position-position coupling\textit{) }by a spring whose elastic constant is $%
k$. If we define $\mathbf{q}_{i}$ as the displacement of mass $m_{i}$ from
the stable equilibrium position, and $\mathbf{p}_{i}$ as its conjugate
momentum, the Hamiltonian for this system will be 
\begin{equation*}
H=\frac{\mathbf{p}_{1}^{2}}{2m_{1}}+\frac{\mathbf{p}_{2}^{2}}{2m_{2}}+\frac{%
k_{1}}{2}\mathbf{q}_{1}^{2}+\frac{k_{2}}{2}\mathbf{q}_{2}^{2}+\frac{k}{2}%
\left( \mathbf{q}_{2}-\mathbf{q}_{1}\right) ^{2},
\end{equation*}
which may be put in the form of equation (\ref{hamitoniano1}) by setting 
\begin{equation*}
\alpha _{i}\equiv \left( m_{i}\left( k_{i}+k\right) \right) ^{\frac{1}{4}},%
\text{ \ \ \ }\omega _{i}\equiv \sqrt{\frac{k_{i}+k}{m_{i}}},\text{ \ \ \ }%
g\equiv -\frac{k}{\alpha _{1}\alpha _{2}},
\end{equation*}
\begin{equation*}
q_{i}\equiv \alpha _{i}\mathbf{q}_{i}\text{ \ \ \ and \ \ \ }p_{i}\equiv
\alpha _{i}^{-1}\mathbf{p}_{i}.
\end{equation*}

The Hamiltonian (\ref{hamitoniano1}) may be written as a product of
matrices: 
\begin{equation*}
H=\mathbf{x}^{T}\cdot \mathbf{H\cdot x,}
\end{equation*}
where 
\begin{equation*}
\mathbf{H\equiv }\frac{1}{2}\left( 
\begin{array}{cccc}
\omega _{1} & g & 0 & 0 \\ 
g & \omega _{2} & 0 & 0 \\ 
0 & 0 & \omega _{1} & 0 \\ 
0 & 0 & 0 & \omega _{2}
\end{array}
\right) ,
\end{equation*}
\begin{equation*}
\mathbf{x}^{T}\equiv \left( 
\begin{array}{cccc}
q_{1} & q_{2} & p_{1} & p_{2}
\end{array}
\right) .
\end{equation*}
Consider a variables transformation determined by the matrix $\mathbf{M}$
(whose inverse exist). In order to preserve Hamilton's equations, $\mathbf{M}
$ must obey the symplectic condition \cite{art9}: 
\begin{equation*}
\mathbf{M\cdot J\cdot M}^{T}=\mathbf{J,}
\end{equation*}
where 
\begin{equation*}
\mathbf{J\equiv }\left( 
\begin{array}{cc}
\mathbf{0} & \mathbf{I} \\ 
-\mathbf{I} & \mathbf{0}
\end{array}
\right) ,
\end{equation*}
$\mathbf{0}$ is a $N\times N$ null matrix and $\mathbf{I}$ is a $N\times N$
identity matrix (here $N=2$, in Section 2.2 $N$ will be generic). In this
new variables, the Hamiltonian is given by 
\begin{equation*}
H=\mathbf{\bar{x}}^{T}\cdot \mathbf{\bar{H}}\cdot \mathbf{\bar{x},}
\end{equation*}
where 
\begin{equation*}
\mathbf{\bar{x}\equiv \mathbf{M}\cdot x}\text{ \ \ \ and \ \ }\mathbf{\bar{H}%
}\equiv \left( \mathbf{M}^{T}\right) ^{-1}\cdot \mathbf{H\cdot \mathbf{M}}%
^{-1}.
\end{equation*}
It will be very easy to find the equations of motion for the system if $%
\mathbf{M}$ is such that $\mathbf{\bar{H}}$ has the form 
\begin{equation}
\mathbf{\bar{H}}=\frac{1}{2}\left( 
\begin{array}{cccc}
\Omega _{+} & 0 & 0 & 0 \\ 
0 & \Omega _{-} & 0 & 0 \\ 
0 & 0 & \Omega _{+} & 0 \\ 
0 & 0 & 0 & \Omega _{-}
\end{array}
\right) .  \label{diagonalizada}
\end{equation}
We will search this $\mathbf{M}$ by performing sections in the phase space
that shall guide us in a sequence of canonical transformations. The matrix $%
\mathbf{M}$ is concerned with the whole transformations carried out in the
sequence, and it will be the product of the matrices of each individual
transformation. Since the following individual transformations obey the
symplectic condition, $\mathbf{M}$ obeys too.

\subsubsection{\textbf{The Transformations}}

\paragraph{\textbf{SQUEEZING IN }$p_{1}\times p_{2}$\textbf{:}}

\begin{figure}[!ht]
\includegraphics[width=16cm]{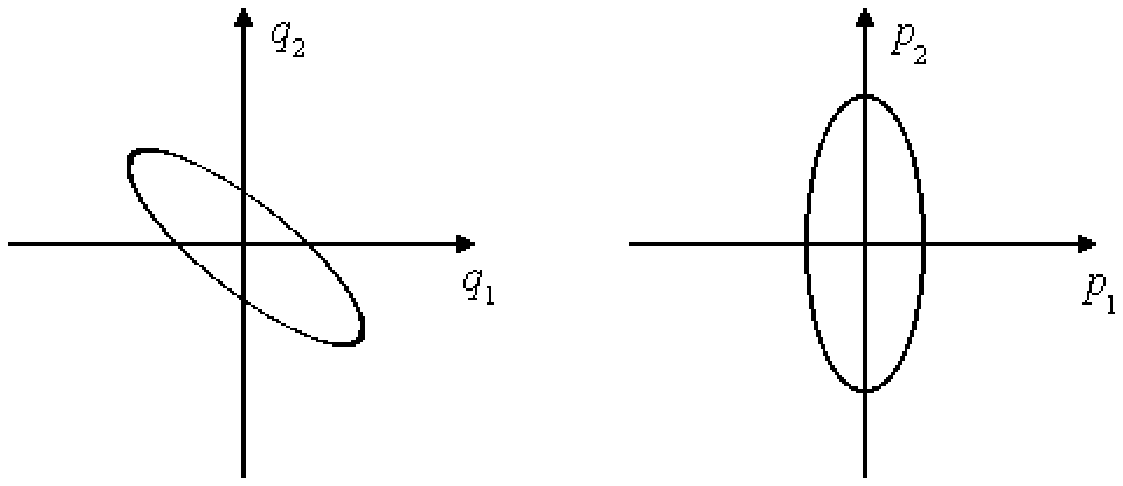}
\vspace{-8cm}
\caption{Sections in the phase space: 
$q_{1}\times q_{2}$ and $p_{1}\times p_{2}$.}
\label{o1}
\end{figure}

Fig. (\ref{o1}) exhibits the shapes of two sections in the phase space.

The ellipse in the $q_{1}\times q_{2}$ plot is inclined due to the term $%
gq_{1}q_{2}$, which could be eliminated by a variables transformation
equivalent to a rotation in the plane $q_{1}\times q_{2}.$ However, the
symplectic condition obliges us to make the rotation in the plane $p_{1}\times
p_{2}$ too, and it would produce a term proportional to $p_{1}p_{2}$. The
solution is to perform a squeezing in the plane $p_{1}\times p_{2}$ first,
turning the ellipse into a circle. Thus the rotation we shall make later
will not cause a new coupling term. \ 

The matrix that carry out this squeezing is 
\begin{equation*}
\mathbf{M}_{1}\equiv \left( 
\begin{array}{cccc}
\alpha & 0 & 0 & 0 \\ 
0 & \alpha ^{-1} & 0 & 0 \\ 
0 & 0 & \alpha ^{-1} & 0 \\ 
0 & 0 & 0 & \alpha
\end{array}
\right) ,
\end{equation*}
where 
\begin{equation*}
\alpha \equiv \left( \frac{\omega _{2}}{\omega _{1}}\right) ^{\frac{1}{4}}.
\end{equation*}
Defining 
\begin{equation*}
\omega \equiv \sqrt{\omega _{1}\omega _{2}},\text{ \ \ \ }\Omega _{1}\equiv 
\sqrt{\frac{\omega _{1}^{3}}{\omega _{2}}}\text{\ \ \ \ and \ \ \ }\Omega
_{2}\equiv \sqrt{\frac{\omega _{2}^{3}}{\omega _{1}}},
\end{equation*}
we may write

\begin{equation*}
\mathbf{H}^{(1)}\mathbf{\equiv }\left( \mathbf{M}_{1}^{T}\right) ^{-1}%
\mathbf{\cdot H\cdot M}_{1}^{-1}=\frac{1}{2}\left( 
\begin{array}{cccc}
\Omega _{1} & g & 0 & 0 \\ 
g & \Omega _{2} & 0 & 0 \\ 
0 & 0 & \omega & 0 \\ 
0 & 0 & 0 & \omega
\end{array}
\right) .
\end{equation*}
The new variables are given by 
\begin{equation*}
\left( 
\begin{array}{c}
q_{1}^{(1)} \\ 
q_{2}^{(1)} \\ 
p_{1}^{(1)} \\ 
p_{2}^{(1)}
\end{array}
\right) \equiv \mathbf{x}^{(1)}\equiv \mathbf{M}_{1\cdot }\mathbf{x},
\end{equation*}
and the Hamiltonian may be represented in the form 
\begin{equation*}
H=\left( \mathbf{x}^{(1)}\right) ^{T}\cdot \mathbf{H}^{(1)}\mathbf{\cdot x}%
^{(1)}\mathbf{\ .}
\end{equation*}

\begin{figure}[!ht]
\includegraphics[width=16cm]{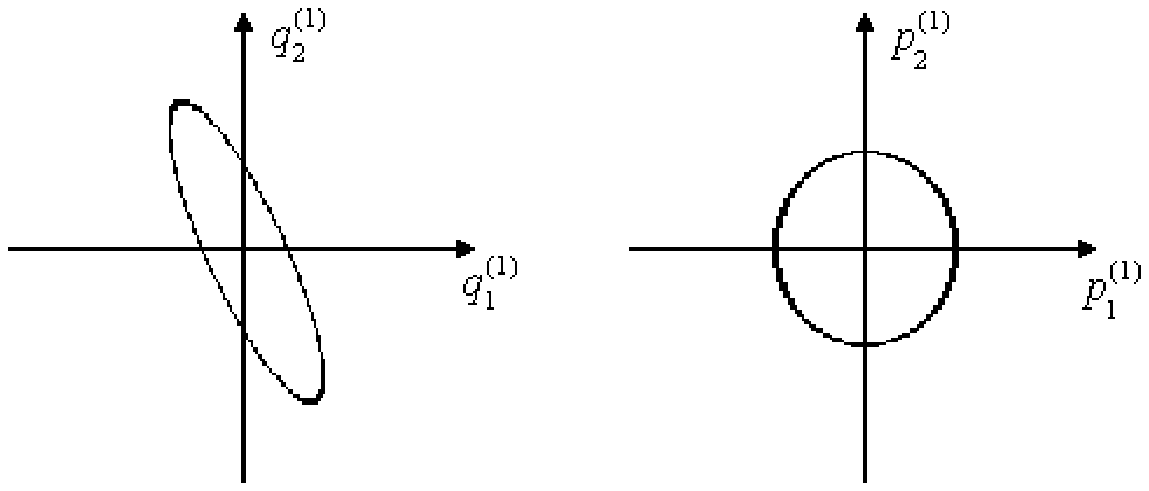}
\vspace{-8cm}
\caption{Sections in the phase space: 
$q_{1}^{\left( 1\right) }\times q_{2}^{\left( 1\right) }$ and $p_{1}^{\left(
1\right) }\times p_{2}^{\left( 1\right) }$.}
\label{o2}
\end{figure}

The sections will be like the ones in Fig. (\ref{o2}).

\paragraph{\textbf{ROTATION IN }$q_{1}^{(1)}\times q_{2}^{(1)}$\textbf{:}}

Now the rotation will not create new nondiagonal terms. The transformation
matrix is 
\begin{equation*}
\mathbf{M}_{2}\equiv \left( 
\begin{array}{cccc}
\cos \varphi & \sin \varphi & 0 & 0 \\ 
-\sin \varphi & \cos \varphi & 0 & 0 \\ 
0 & 0 & \cos \varphi & \sin \varphi \\ 
0 & 0 & -\sin \varphi & \cos \varphi
\end{array}
\right) ,
\end{equation*}
and we define 
\begin{eqnarray*}
\mathbf{H}^{(2)} &\equiv &\left( \mathbf{M}_{2}^{T}\right) ^{-1}\mathbf{%
\cdot H}^{(1)}\mathbf{\cdot M}_{2}^{-1} \\
&\equiv &\frac{1}{2}\left( 
\begin{array}{cc}
\mathbf{H}_{11}^{(2)} & \mathbf{H}_{12}^{(2)} \\ 
\mathbf{H}_{21}^{(2)} & \mathbf{H}_{22}^{(2)}
\end{array}
\right) ,
\end{eqnarray*}
where 
\begin{equation*}
\mathbf{H}_{11}^{(2)}=\left( 
\begin{array}{cc}
\Omega _{1}\cos ^{2}\varphi +\Omega _{2}\sin ^{2}\varphi +2g\sin \varphi
\cos \varphi & \sin \varphi \cos \varphi \left( \Omega _{2}-\Omega
_{1}\right) +g\left( \cos ^{2}\varphi -\sin ^{2}\varphi \right) \\ 
\sin \varphi \cos \varphi \left( \Omega _{2}-\Omega _{1}\right) +g\left(
\cos ^{2}\varphi -\sin ^{2}\varphi \right) & \Omega _{1}\sin ^{2}\varphi
+\Omega _{2}\cos ^{2}\varphi -2g\sin \varphi \cos \varphi
\end{array}
\right) ,
\end{equation*}
\begin{equation*}
\mathbf{H}_{12}^{(2)}=\mathbf{H}_{21}^{(2)}=\left( 
\begin{array}{cc}
0 & 0 \\ 
0 & 0
\end{array}
\right) ,\text{ \ \ \ }\mathbf{H}_{22}^{(2)}=\left( 
\begin{array}{cc}
\omega & 0 \\ 
0 & \omega
\end{array}
\right) .
\end{equation*}
We now choose 
\begin{equation*}
\cos \varphi =\sqrt{\frac{1}{2}\left( 1+\frac{\Omega _{1}-\Omega _{2}}{\sqrt{%
\left( \Omega _{1}-\Omega _{2}\right) ^{2}+4g^{2}}}\right) },
\end{equation*}
\begin{equation*}
\sin \varphi =-\sqrt{\frac{1}{2}\left( 1-\frac{\Omega _{1}-\Omega _{2}}{%
\sqrt{\left( \Omega _{1}-\Omega _{2}\right) ^{2}+4g^{2}}}\right) },
\end{equation*}
in order to make $\mathbf{H}^{(2)}$ diagonal. Thus 
\begin{equation*}
\mathbf{H}^{(2)}=\frac{1}{2}\left( 
\begin{array}{cccc}
\omega _{+} & 0 & 0 & 0 \\ 
0 & \omega _{-} & 0 & 0 \\ 
0 & 0 & \omega & 0 \\ 
0 & 0 & 0 & \omega
\end{array}
\right) ,
\end{equation*}
where 
\begin{equation*}
\omega _{+}\equiv \frac{1}{2}\left( \left( \Omega _{1}+\Omega _{2}\right) +%
\sqrt{\left( \Omega _{1}-\Omega _{2}\right) ^{2}+4g^{2}}\right) ,
\end{equation*}
\begin{equation*}
\omega _{-}\equiv \frac{1}{2}\left( \left( \Omega _{1}+\Omega _{2}\right) -%
\sqrt{\left( \Omega _{1}-\Omega _{2}\right) ^{2}+4g^{2}}\right) .
\end{equation*}
Defining the new variables as

\begin{equation*}
\left( 
\begin{array}{c}
q_{1}^{(2)} \\ 
q_{2}^{(2)} \\ 
p_{1}^{(2)} \\ 
p_{2}^{(2)}
\end{array}
\right) \equiv \mathbf{x}^{(2)}\equiv \mathbf{M}_{2}\cdot \mathbf{x}^{(1)},
\end{equation*}
we may write the Hamiltonian in the form 
\begin{equation*}
H=\left( \mathbf{x}^{(2)}\right) ^{T}\cdot \mathbf{H}^{(2)}\mathbf{\cdot x}%
^{(2)}\mathbf{.}
\end{equation*}

\begin{figure}[!ht]
\includegraphics[width=16cm]{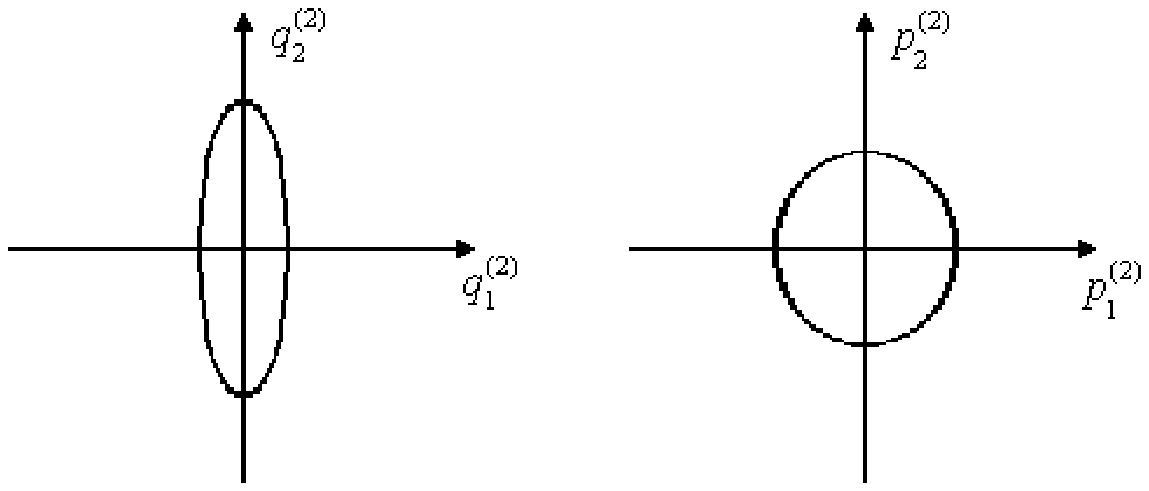}
\vspace{-8cm}
\caption{Sections in the phase space: 
$q_{1}^{\left( 2\right) }\times q_{2}^{\left( 2\right) }$ and $p_{1}^{\left(
2\right) }\times p_{2}^{\left( 2\right) }$.}
\label{o3}
\end{figure}

For these rotated variables, the sections will be like the ones in Fig (\ref
{o3}).

\paragraph{\textbf{SQUEEZING IN }$q_{1}^{(2)}\times p_{1}^{(2)}$\textbf{\
AND IN }$q_{2}^{(2)}\times p_{2}^{(2)}$\textbf{:}}

\begin{figure}[!ht]
\includegraphics[width=16cm]{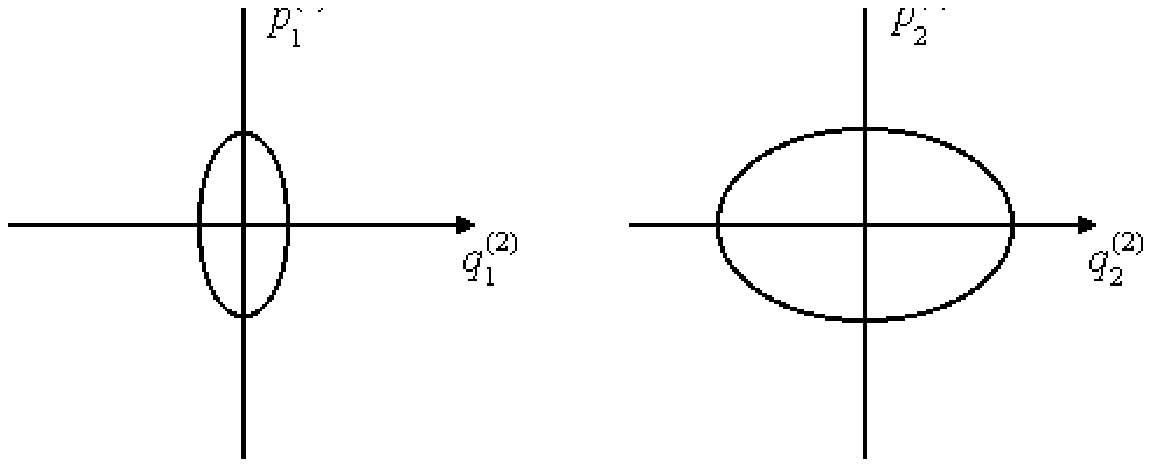}
\vspace{-8cm}
\caption{Sections in the phase space: 
$q_{1}^{\left( 2\right) }\times p_{1}^{\left( 2\right) }$ and $q_{2}^{\left(
2\right) }\times p_{2}^{\left( 2\right) }$.}
\label{o4}
\end{figure}

The matrix $\mathbf{H}^{(2)}$ is not in the final form (equation (\ref
{diagonalizada})), because the coefficient of each coordinate is not equal
to the coefficient of its conjugate momentum, as may be seen in Fig (\ref{o4}%
).

The following transformation does the adjustment: 
\begin{equation*}
\mathbf{M}_{3}\equiv \left( 
\begin{array}{cccc}
\alpha _{+} & 0 & 0 & 0 \\ 
0 & \alpha _{-} & 0 & 0 \\ 
0 & 0 & \alpha _{+}^{-1} & 0 \\ 
0 & 0 & 0 & \alpha _{-}^{-1}
\end{array}
\right) ,
\end{equation*}
where 
\begin{equation*}
\alpha _{\pm }\equiv \left( \frac{\omega _{\pm }}{\omega }\right) ^{1/4}.
\end{equation*}
We find the matrix $\mathbf{\bar{H}}$ (equation (\ref{diagonalizada})) by
defining 
\begin{eqnarray*}
\mathbf{\bar{H}} &\equiv &\left( \mathbf{M}_{3}^{T}\right) ^{-1}\cdot 
\mathbf{H}^{(2)}\mathbf{\cdot \mathbf{M}}_{3}^{-1} \\
&=&\frac{1}{2}\left( 
\begin{array}{cccc}
\Omega _{+} & 0 & 0 & 0 \\ 
0 & \Omega _{-} & 0 & 0 \\ 
0 & 0 & \Omega _{+} & 0 \\ 
0 & 0 & 0 & \Omega _{-}
\end{array}
\right) ,
\end{eqnarray*}
where 
\begin{equation*}
\Omega _{\pm }\equiv \sqrt{\omega \omega _{\pm }}.
\end{equation*}
The matrix which corresponds to the set of transformations carried out is 
\begin{equation*}
\mathbf{M}\equiv \mathbf{M}_{3}\cdot \mathbf{M}_{2}\cdot \mathbf{M}_{1}.
\end{equation*}
Thus we may write

\begin{equation*}
\mathbf{\bar{H}}=\left( \mathbf{M}^{T}\right) ^{-1}\cdot \mathbf{H\cdot 
\mathbf{M}}^{-1}\mathbf{,}
\end{equation*}
and the new variables are 
\begin{equation*}
\left( 
\begin{array}{c}
\bar{q}_{1} \\ 
\bar{q}_{2} \\ 
\bar{p}_{1} \\ 
\bar{p}_{2}
\end{array}
\right) \equiv \mathbf{\bar{x}\equiv M}\cdot \mathbf{x}.
\end{equation*}
The expression for the Hamiltonian using $\mathbf{\bar{x}}$ is 
\begin{equation*}
H=\mathbf{\bar{x}}^{T}\cdot \mathbf{\bar{H}\cdot \bar{x}.}
\end{equation*}

\begin{figure}[!ht]
\includegraphics[width=16cm]{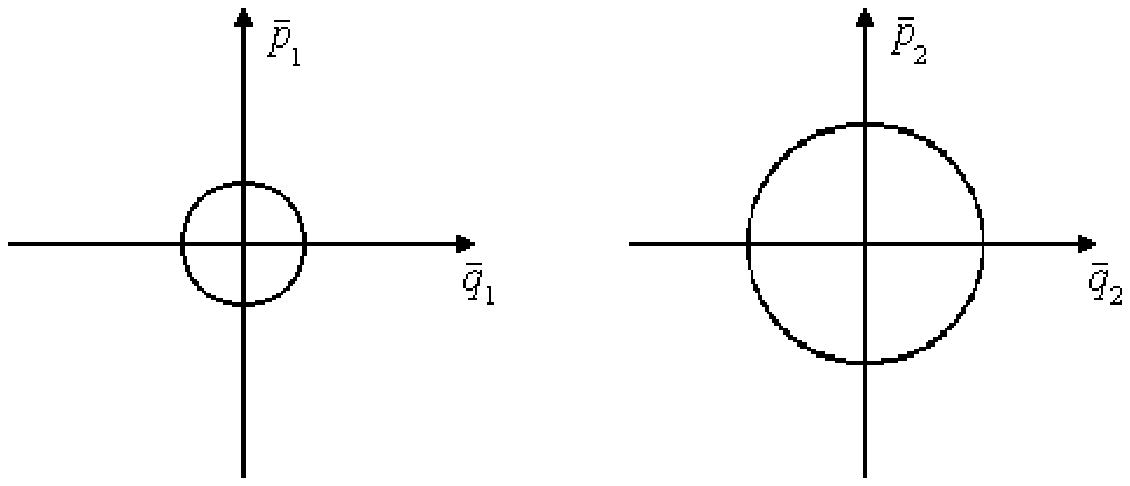}
\vspace{-8cm}
\caption{Sections in the phase space: 
$\bar{q}_{1}\times \bar{p}_{1}$ and $\bar{q}_{2}\times \bar{p}_{2}$.}
\label{o5}
\end{figure}

Fig. (\ref{o5}) shows the equality of the coefficient of each coordinate and
its conjugate momentum.

The parameters $\Omega _{\pm }$ are given in terms of the original
parameters by 
\begin{equation*}
\Omega _{\pm }=\frac{1}{\sqrt{2}}\sqrt{\omega _{1}^{2}+\omega _{2}^{2}\pm 
\sqrt{\left( \omega _{1}^{2}-\omega _{2}^{2}\right) ^{2}+4g^{2}\omega
_{1}\omega _{2}}}.
\end{equation*}
The constant $\Omega _{+}$ is always real, since $\omega _{1}$ and $\omega
_{2}$ are positive real numbers. The constant $\Omega _{-}$ will be real
when 
\begin{equation*}
\omega _{1}\omega _{2}\geq g^{2}.
\end{equation*}
If we are dealing with a spring-mass system like the one in Fig. (\ref{OA}),
this condition is always satisfied, since it corresponds to 
\begin{equation*}
\left( k_{1}+k\right) \left( k_{2}+k\right) \geq k^{2}.
\end{equation*}

\subsubsection{The Equations of Motion:}

Since $\mathbf{M}$ refers to a canonical transformation, the Hamilton's
equations are preserved. Thus we may write them, in the matricial form 
\cite{art9},
for the variables in $\mathbf{\bar{x}}$: 
\begin{equation*}
\overset{\bullet }{\mathbf{\bar{x}}}=\mathbf{J\cdot }\frac{\partial }{%
\partial \mathbf{\bar{x}}}H=\left( \mathbf{J\cdot }\frac{\partial }{\partial 
\mathbf{\bar{x}}}\right) \cdot \left( \mathbf{\bar{x}}^{T}\cdot \mathbf{\bar{%
H}\cdot \bar{x}}\right) \mathbf{,}
\end{equation*}
where 
\begin{equation*}
\left( \frac{\partial }{\partial \mathbf{\bar{x}}}\right) ^{T}\equiv \left( 
\begin{array}{cccc}
\frac{\partial }{\partial \bar{q}}_{1} & \frac{\partial }{\partial \bar{q}%
_{2}} & \frac{\partial }{\partial \bar{p}_{1}} & \frac{\partial }{\partial 
\bar{p}_{2}}
\end{array}
\right) .
\end{equation*}
It yields 
\begin{equation*}
\left( 
\begin{array}{c}
\frac{\partial }{\partial t}\bar{q}_{1} \\ 
\frac{\partial }{\partial t}\bar{q}_{2} \\ 
\frac{\partial }{\partial t}\bar{p}_{1} \\ 
\frac{\partial }{\partial t}\bar{p}_{2}
\end{array}
\right) =\left( 
\begin{array}{c}
\Omega _{+}\bar{p}_{1} \\ 
\Omega _{-}\bar{p}_{2} \\ 
-\Omega _{+}\bar{q}_{1} \\ 
-\Omega _{-}\bar{q}_{2}
\end{array}
\right) ,
\end{equation*}
whose solution is 
\begin{equation*}
\left( 
\begin{array}{c}
\bar{q}_{1}(t) \\ 
\bar{q}_{2}(t) \\ 
\bar{p}_{1}(t) \\ 
\bar{p}_{2}(t)
\end{array}
\right) =\left( 
\begin{array}{cccc}
\cos (\Omega _{+}t) & 0 & \sin (\Omega _{+}t) & 0 \\ 
0 & \cos (\Omega _{-}t) & 0 & \sin (\Omega _{-}t) \\ 
-\sin (\Omega _{+}t) & 0 & \cos (\Omega _{+}t) & 0 \\ 
0 & -\sin (\Omega _{-}t) & 0 & \cos (\Omega _{-}t)
\end{array}
\right) \cdot \left( 
\begin{array}{c}
\bar{q}_{1}(0) \\ 
\bar{q}_{2}(0) \\ 
\bar{p}_{1}(0) \\ 
\bar{p}_{2}(0)
\end{array}
\right) .
\end{equation*}
It is clear in the equation above that the frequencies of the normal modes
are $\Omega _{+}$ and $\Omega _{-}$.

Defining 
\begin{equation*}
\mathbf{N}\equiv \left( 
\begin{array}{cccc}
\cos (\Omega _{+}t) & 0 & \sin (\Omega _{+}t) & 0 \\ 
0 & \cos (\Omega _{-}t) & 0 & \sin (\Omega _{-}t) \\ 
-\sin (\Omega _{+}t) & 0 & \cos (\Omega _{+}t) & 0 \\ 
0 & -\sin (\Omega _{-}t) & 0 & \cos (\Omega _{-}t)
\end{array}
\right) ,
\end{equation*}
we may write the equations of motion for the original variables: 
\begin{equation*}
\mathbf{x}(t)=\mathbf{M}^{-1}\mathbf{\cdot \bar{x}}(t)=\mathbf{M}^{-1}%
\mathbf{\cdot N\cdot \bar{x}}(0)=\mathbf{M}^{-1}\cdot \mathbf{N\cdot M}\cdot 
\mathbf{x}(0).
\end{equation*}

\bigskip

\subsection{$N$ Coupled Oscillators:}

Consider now one oscillator linearly coupled to $N-1$ other oscillators. Of
course the system we just solved is a particular case of this system. Thus
the transformations we found there are useful to suggest the form of the
transformations to be effected here.

The $N$ coupled oscillators' Hamiltonian is 
\begin{equation*}
H=\frac{1}{2}\overset{N}{\underset{i=1}{\sum }}\omega _{i,i}\left(
p_{i}^{2}+q_{i}^{2}\right) +\overset{N}{\underset{i=2}{\sum }}\omega _{1,i}\
q_{1}q_{i},
\end{equation*}
where $q_{i}$ and $p_{i}$ are canonical conjugate variables, all the $\omega
_{i,j}$ are real numbers, $\omega _{i,i}>0$ and $\omega _{i,j}\leqslant 0$
for $i\neq j$. This Hamiltonian may be written in a matricial form: 
\begin{equation*}
H=\frac{1}{2}\left( 
\begin{array}{cc}
\mathbf{Q}^{T} & \mathbf{P}^{T}
\end{array}
\right) \left( 
\begin{array}{cc}
\mathbf{H}_{\mathbf{Q}} & \mathbf{0} \\ 
\mathbf{0} & \mathbf{H}_{\mathbf{P}}
\end{array}
\right) \left( 
\begin{array}{c}
\mathbf{Q} \\ 
\mathbf{P}
\end{array}
\right) ,
\end{equation*}
where 
\begin{eqnarray*}
\mathbf{H}_{\mathbf{Q}} &\equiv &\left( 
\begin{array}{ccccc}
\omega _{1,1} & \omega _{1,2} & \cdots & \omega _{1,N-i} & \omega _{1,N} \\ 
\omega _{1,2} & \omega _{2,2} & 0 & \cdots & 0 \\ 
\vdots & 0 & \ddots & \ddots & \vdots \\ 
\omega _{1,N-i} & \vdots & \ddots & \omega _{N-1,N-1} & 0 \\ 
\omega _{1,N} & 0 & \cdots & 0 & \omega _{N,N}
\end{array}
\right) , \\
\mathbf{H}_{\mathbf{P}} &\equiv &\left( 
\begin{array}{ccccc}
\omega _{1,1} & 0 & \cdots & 0 & 0 \\ 
0 & \omega _{2,2} & 0 & \cdots & 0 \\ 
\vdots & 0 & \ddots & \ddots & \vdots \\ 
0 & \vdots & \ddots & \omega _{N-1,N-1} & 0 \\ 
0 & 0 & \cdots & 0 & \omega _{N,N}
\end{array}
\right) , \\
\mathbf{Q}^{T} &\equiv &\left( 
\begin{array}{ccccc}
q_{1} & q_{2} & \cdots & q_{N-1} & q_{N}
\end{array}
\right) , \\
\mathbf{P}^{T} &\equiv &\left( 
\begin{array}{ccccc}
p_{1} & p_{2} & \cdots & p_{N-1} & p_{N}
\end{array}
\right) .
\end{eqnarray*}

The aim of the following transformations is to diagonalize the matrix 
\begin{equation*}
\left( 
\begin{array}{cc}
\mathbf{H}_{\mathbf{Q}} & \mathbf{0} \\ 
\mathbf{0} & \mathbf{H}_{\mathbf{P}}
\end{array}
\right) .
\end{equation*}
All of them are canonical, as may be verified using the symplectic condition,
and reduce to the transformations found in Section 2.1 when $N=2$.

\subsubsection{\textbf{The Transformations}}

\paragraph{\textbf{SQUEEZINGS IN }$\mathbf{P}$\textbf{:}}

Regard 
\begin{equation*}
\mathbf{M}_{\mathbf{S}}\equiv \mathbf{S}_{1}\cdot \mathbf{S}_{2}\cdot
...\cdot \mathbf{S}_{N-1}\cdot \mathbf{S}_{N},
\end{equation*}
where 
\begin{equation*}
\mathbf{S}_{i}\equiv \left( \omega _{i,i}\right) ^{\frac{1}{4}}\left( 
\begin{array}{ccccc}
1 & 0 & \cdots & 0 & 0 \\ 
0 & \ddots & \ddots & \vdots & 0 \\ 
\vdots & \ddots & \left( \omega _{i,i}\right) ^{-\frac{1}{2}} & 0 & \vdots
\\ 
0 & \cdots & 0 & \ddots & 0 \\ 
0 & 0 & \cdots & 0 & 1
\end{array}
\right)
\end{equation*}
($\left[ \mathbf{S}_{i}\right] _{i,i}=\left( \omega _{i,i}\right) ^{\frac{1}{%
4}}\left( \omega _{i,i}\right) ^{-\frac{1}{2}}$, $\left[ \mathbf{S}_{i}%
\right] _{j,k}=\left( \omega _{i,i}\right) ^{\frac{1}{4}}\delta _{j,k}$ for $%
j\neq i$). The variables transformation to be effected is 
\begin{equation*}
\left( 
\begin{array}{c}
\mathbf{Q}_{\mathbf{S}} \\ 
\mathbf{P}_{\mathbf{S}}
\end{array}
\right) \equiv \left( 
\begin{array}{cc}
\mathbf{M}_{\mathbf{S}} & \mathbf{0} \\ 
\mathbf{0} & \mathbf{M}_{\mathbf{S}}^{-1}
\end{array}
\right) \left( 
\begin{array}{c}
\mathbf{Q} \\ 
\mathbf{P}
\end{array}
\right) ,
\end{equation*}
and the Hamiltonian may be written as 
\begin{equation*}
H=\frac{1}{2}\left( 
\begin{array}{cc}
\mathbf{Q}_{\mathbf{S}}^{T} & \mathbf{P}_{\mathbf{S}}^{T}
\end{array}
\right) \left( 
\begin{array}{cc}
\mathbf{H}_{\mathbf{Q}_{\mathbf{S}}} & \mathbf{0} \\ 
\mathbf{0} & \mathbf{H}_{\mathbf{P}_{\mathbf{S}}}
\end{array}
\right) \left( 
\begin{array}{c}
\mathbf{Q}_{\mathbf{S}} \\ 
\mathbf{P}_{\mathbf{S}}
\end{array}
\right) ,
\end{equation*}
where 
\begin{eqnarray*}
\left( 
\begin{array}{cc}
\mathbf{H}_{\mathbf{Q}_{\mathbf{S}}} & \mathbf{0} \\ 
\mathbf{0} & \mathbf{H}_{\mathbf{P}_{\mathbf{S}}}
\end{array}
\right) &\equiv &\left( \left( 
\begin{array}{cc}
\mathbf{M}_{\mathbf{S}} & \mathbf{0} \\ 
\mathbf{0} & \mathbf{M}_{\mathbf{S}}^{-1}
\end{array}
\right) ^{T}\right) ^{-1}\left( 
\begin{array}{cc}
\mathbf{H}_{\mathbf{Q}} & \mathbf{0} \\ 
\mathbf{0} & \mathbf{H}_{\mathbf{P}}
\end{array}
\right) \left( 
\begin{array}{cc}
\mathbf{M}_{\mathbf{S}} & \mathbf{0} \\ 
\mathbf{0} & \mathbf{M}_{\mathbf{S}}^{-1}
\end{array}
\right) ^{-1} \\
&=&\left( 
\begin{array}{cc}
\mathbf{M}_{\mathbf{S}}^{-1}\cdot \mathbf{H}_{\mathbf{Q}}\cdot \mathbf{M}_{%
\mathbf{S}}^{-1} & \mathbf{0} \\ 
\mathbf{0} & \mathbf{M}_{\mathbf{S}}\cdot \mathbf{H}_{\mathbf{P}}\cdot 
\mathbf{M}_{\mathbf{S}}
\end{array}
\right) .
\end{eqnarray*}
The matrices $\mathbf{H}_{\mathbf{Q}_{\mathbf{S}}}$ and $\mathbf{H}_{\mathbf{%
P}_{\mathbf{S}}}$ are given by 
\begin{eqnarray}
\mathbf{H}_{\mathbf{Q}_{\mathbf{S}}} &=&\left( 
\begin{array}{ccccc}
g_{1,1} & g_{1,2} & \cdots & g_{1,N-1} & g_{1,N} \\ 
g_{1,2} & g_{2,2} & 0 & \cdots & 0 \\ 
\vdots & 0 & \ddots & \ddots & \vdots \\ 
g_{1,N-1} & \vdots & \ddots & g_{N-1,N-1} & 0 \\ 
g_{1,N} & 0 & \cdots & 0 & g_{N,N}
\end{array}
\right) ,  \label{HQS} \\
\mathbf{H}_{\mathbf{P}_{\mathbf{S}}} &=&\left( 
\begin{array}{ccccc}
G & 0 & \cdots & 0 & 0 \\ 
0 & G & \ddots & \vdots & 0 \\ 
\vdots & \ddots & \ddots & 0 & \vdots \\ 
0 & \cdots & 0 & G & 0 \\ 
0 & 0 & \cdots & 0 & G
\end{array}
\right) ,  \notag
\end{eqnarray}
with 
\begin{eqnarray*}
G &=&\overset{N}{\underset{i=1}{\prod }}\sqrt{\omega _{i,i}}, \\
g_{i,j} &=&\frac{\omega _{i,j}\sqrt{\omega _{i,i}\omega _{j,j}}}{G}.
\end{eqnarray*}

This transformation corresponds to $N$ squeezings, each one involving $q_{i}$
and $p_{i}$. Notice that $G$ and $g_{i,j}$ are real numbers ($G>0$, $%
g_{i,i}>0$ and $g_{i,j}\leqslant 0$ for $i\neq j$).

\paragraph{\textbf{ROTATIONS IN }$\mathbf{Q}_{\mathbf{S}}$\textbf{:}}

Since $\mathbf{H}_{\mathbf{Q}_{\mathbf{S}}}$ is a real symmetric matrix, it
has real eigenvalues and may be diagonalized by a real orthogonal matrix $%
\mathbf{M}_{\mathbf{R}}$. Thus we define 
\begin{eqnarray}
\mathbf{H}_{\mathbf{Q}_{\mathbf{R}}} &\equiv &\mathbf{M}_{\mathbf{R}}\cdot 
\mathbf{H}_{\mathbf{Q}_{\mathbf{S}}}\cdot \mathbf{M}_{\mathbf{R}}^{T}
\label{HQR} \\
&\equiv &\left( 
\begin{array}{ccccc}
\lambda _{1,1} & 0 & \cdots & 0 & 0 \\ 
0 & \lambda _{2,2} & 0 & \cdots & 0 \\ 
\vdots & 0 & \ddots & \ddots & \vdots \\ 
0 & \vdots & \ddots & \lambda _{N-1,N-1} & 0 \\ 
0 & 0 & \cdots & 0 & \lambda _{N,N}
\end{array}
\right) ,  \notag
\end{eqnarray}
where the $\lambda _{i,i}$ are real numbers and 
\begin{equation}
\mathbf{M}_{\mathbf{R}}\cdot \mathbf{M}_{\mathbf{R}}^{T}=\mathbf{I},
\label{ortonormalidade}
\end{equation}
($\mathbf{I}$ is the $N\times N$ identity matrix). To find the $\lambda
_{i,i}$ may be non trivial, since, to this end, it would be necessary to
solve a $N$ degree polynomial. We do not have a general expression for $%
\mathbf{M}_{\mathbf{R}}$, but we show in the Appendix that it is possible to
write $\mathbf{M}_{\mathbf{R}}$ as the product 
\begin{equation}
\mathbf{M}_{\mathbf{R}}=\mathbf{R}_{\mathbf{N-1}}\cdot \mathbf{R}_{\mathbf{%
N-2}}\cdot ...\cdot \mathbf{R}_{\mathbf{2}}\cdot \mathbf{R}_{\mathbf{1}},
\label{MR=RN-1...R1}
\end{equation}
where 
\begin{equation*}
\mathbf{R}_{\mathbf{i}}\equiv \mathbf{R}_{i,N}\cdot \mathbf{R}_{i,N-1}\cdot
...\cdot \mathbf{R}_{i,i+2}\cdot \mathbf{R}_{i,i+1},
\end{equation*}
and the $\mathbf{R}_{i,j}$ are matrices whose elements $\left[ \mathbf{R}%
_{i,j}\right] _{k,l}$ are given by 
\begin{eqnarray*}
\left[ \mathbf{R}_{i,j}\right] _{i,i} &=&\left[ \mathbf{R}_{i,j}\right]
_{j,j}=\cos \alpha _{i,j}, \\
\left[ \mathbf{R}_{i,j}\right] _{i,j} &=&\sin \alpha _{i,j}, \\
\left[ \mathbf{R}_{i,j}\right] _{j,i} &=&-\sin \alpha _{i,j}, \\
\left[ \mathbf{R}_{i,j}\right] _{k,k} &=&1\text{ for }k\neq i\text{ and }%
k\neq j, \\
\left[ \mathbf{R}_{i,j}\right] _{k,l} &=&0\text{ for the other cases.}
\end{eqnarray*}
We may interpret $\mathbf{R}_{i,j}$ as rotation in a $N$ dimensional space,
and $\mathbf{M}_{\mathbf{R}}$ \ as a sequence of such rotations.

An intuitive view of the $\mathbf{M}_{\mathbf{R}}$ decomposition may be
achieved. Equation (\ref{ortonormalidade}) permit us to interpret the $%
\mathbf{M}_{\mathbf{R}}$ lines as the coordinates of vectors which form an
orthonormal basis in a $N$ dimensional space. Of course this coordinates are
written in another orthonormal basis, which we call the canonical basis. If
we apply a suitable sequence of rotations (represented by $\mathbf{M}_{%
\mathbf{R}}^{T}$) in the basis represented by $\mathbf{M}_{\mathbf{R}}$, we
can make it coincident to the canonical basis, and it will be represented by
an identity matrix ($\mathbf{M}_{\mathbf{R}}^{T}\cdot \mathbf{M}_{\mathbf{R}%
}=\mathbf{I}$). In the $N=2$ and $N=3$ cases it is not difficult to imagine
the rotations being performed.

The transformation to be effected here is 
\begin{equation*}
\left( 
\begin{array}{c}
\mathbf{Q}_{\mathbf{R}} \\ 
\mathbf{P}_{\mathbf{R}}
\end{array}
\right) \equiv \left( 
\begin{array}{cc}
\mathbf{M}_{\mathbf{R}} & \mathbf{0} \\ 
\mathbf{0} & \mathbf{M}_{\mathbf{R}}
\end{array}
\right) \left( 
\begin{array}{cc}
\mathbf{M}_{\mathbf{S}} & \mathbf{0} \\ 
\mathbf{0} & \mathbf{M}_{\mathbf{S}}^{-1}
\end{array}
\right) \left( 
\begin{array}{c}
\mathbf{Q} \\ 
\mathbf{P}
\end{array}
\right) .
\end{equation*}
We may write the Hamiltonian as 
\begin{equation*}
H=\frac{1}{2}\left( 
\begin{array}{cc}
\mathbf{Q}_{\mathbf{R}}^{T} & \mathbf{P}_{\mathbf{R}}^{T}
\end{array}
\right) \left( 
\begin{array}{cc}
\mathbf{H}_{\mathbf{Q}_{\mathbf{R}}} & \mathbf{0} \\ 
\mathbf{0} & \mathbf{H}_{\mathbf{P}_{\mathbf{R}}}
\end{array}
\right) \left( 
\begin{array}{c}
\mathbf{Q}_{\mathbf{R}} \\ 
\mathbf{P}_{\mathbf{R}}
\end{array}
\right) ,
\end{equation*}
where 
\begin{eqnarray*}
\left( 
\begin{array}{cc}
\mathbf{H}_{\mathbf{Q}_{\mathbf{R}}} & \mathbf{0} \\ 
\mathbf{0} & \mathbf{H}_{\mathbf{P}_{\mathbf{R}}}
\end{array}
\right) &\equiv &\left( \left( 
\begin{array}{cc}
\mathbf{M}_{\mathbf{R}} & \mathbf{0} \\ 
\mathbf{0} & \mathbf{M}_{\mathbf{R}}
\end{array}
\right) ^{T}\right) ^{-1}\left( 
\begin{array}{cc}
\mathbf{H}_{\mathbf{Q}_{\mathbf{S}}} & \mathbf{0} \\ 
\mathbf{0} & \mathbf{H}_{\mathbf{P}_{\mathbf{S}}}
\end{array}
\right) \left( 
\begin{array}{cc}
\mathbf{M}_{\mathbf{R}} & \mathbf{0} \\ 
\mathbf{0} & \mathbf{M}_{\mathbf{R}}
\end{array}
\right) ^{-1}, \\
&=&\left( 
\begin{array}{cc}
\mathbf{M}_{\mathbf{R}}\cdot \mathbf{H}_{\mathbf{Q}_{\mathbf{S}}}\cdot 
\mathbf{M}_{\mathbf{R}}^{T} & \mathbf{0} \\ 
\mathbf{0} & \mathbf{H}_{\mathbf{P}_{\mathbf{S}}}
\end{array}
\right) .
\end{eqnarray*}

\paragraph{SQUEEZINGS IN $\mathbf{Q}_{\mathbf{R}}\times \mathbf{P}_{\mathbf{R%
}}$:}

Consider now 
\begin{equation*}
\mathbf{M}_{\mathbf{T}}\equiv \mathbf{T}_{\mathbf{1}}\cdot \mathbf{T}_{%
\mathbf{2}}\cdot ...\cdot \mathbf{T}_{\mathbf{N-1}}\cdot \mathbf{T}_{\mathbf{%
N}},
\end{equation*}
where 
\begin{equation*}
\mathbf{T}_{\mathbf{i}}\equiv \left( \omega _{i,i}\right) ^{-\frac{1}{8}%
}\left( 
\begin{array}{ccccc}
1 & 0 & \cdots  & 0 & 0 \\ 
0 & \ddots  & \ddots  & \vdots  & 0 \\ 
\vdots  & \ddots  & \left( \lambda _{i,i}\right) ^{\frac{1}{4}} & 0 & \vdots 
\\ 
0 & \cdots  & 0 & \ddots  & 0 \\ 
0 & 0 & \cdots  & 0 & 1
\end{array}
\right) .
\end{equation*}
Defining 
\begin{eqnarray*}
\left( 
\begin{array}{cc}
\mathbf{H}_{\mathbf{Q}_{\mathbf{T}}} & \mathbf{0} \\ 
\mathbf{0} & \mathbf{H}_{\mathbf{P}_{\mathbf{T}}}
\end{array}
\right)  &\equiv &\left( \left( 
\begin{array}{cc}
\mathbf{M}_{\mathbf{T}} & \mathbf{0} \\ 
\mathbf{0} & \mathbf{M}_{\mathbf{T}}^{-1}
\end{array}
\right) ^{T}\right) ^{-1}\left( 
\begin{array}{cc}
\mathbf{H}_{\mathbf{Q}_{\mathbf{R}}} & \mathbf{0} \\ 
\mathbf{0} & \mathbf{H}_{\mathbf{P}_{\mathbf{R}}}
\end{array}
\right) \left( 
\begin{array}{cc}
\mathbf{M}_{\mathbf{T}} & \mathbf{0} \\ 
\mathbf{0} & \mathbf{M}_{\mathbf{T}}^{-1}
\end{array}
\right) ^{-1} \\
&=&\left( 
\begin{array}{cc}
\mathbf{M}_{\mathbf{T}}^{-1}\cdot \mathbf{H}_{\mathbf{Q}_{\mathbf{R}}}\cdot 
\mathbf{M}_{\mathbf{T}}^{-1} & \mathbf{0} \\ 
\mathbf{0} & \mathbf{M}_{\mathbf{T}}\cdot \mathbf{H}_{\mathbf{P}_{\mathbf{R}%
}}\cdot \mathbf{M}_{\mathbf{T}}
\end{array}
\right) ,
\end{eqnarray*}
we have 
\begin{equation*}
\mathbf{H}_{\mathbf{Q}_{\mathbf{T}}}=\mathbf{H}_{\mathbf{P}_{\mathbf{T}%
}}=\left( 
\begin{array}{ccccc}
\Omega _{1,1} & 0 & 0 & 0 & 0 \\ 
0 & \Omega _{2,2} & 0 & 0 & 0 \\ 
0 & 0 & \ddots  & 0 & 0 \\ 
0 & 0 & 0 & \Omega _{N-1,N-1} & 0 \\ 
0 & 0 & 0 & 0 & \Omega _{N,N}
\end{array}
\right) ,
\end{equation*}
where 
\begin{equation*}
\Omega _{i,i}\equiv \left( G\lambda _{i,i}\right) ^{\frac{1}{2}}.
\end{equation*}
Of course $\mathbf{M}_{\mathbf{T}}$ corresponds to $N$ squeezings like the
ones carried out in Section 2.1.

The final variables will be given by 
\begin{equation*}
\left( 
\begin{array}{c}
\mathbf{Q}_{\mathbf{T}} \\ 
\mathbf{P}_{\mathbf{T}}
\end{array}
\right) \equiv \left( 
\begin{array}{cc}
\mathbf{M}_{\mathbf{T}} & \mathbf{0} \\ 
\mathbf{0} & \mathbf{M}_{\mathbf{T}}^{-1}
\end{array}
\right) \left( 
\begin{array}{cc}
\mathbf{M}_{\mathbf{R}} & \mathbf{0} \\ 
\mathbf{0} & \mathbf{M}_{\mathbf{R}}
\end{array}
\right) \left( 
\begin{array}{cc}
\mathbf{M}_{\mathbf{S}} & \mathbf{0} \\ 
\mathbf{0} & \mathbf{M}_{\mathbf{S}}^{-1}
\end{array}
\right) \left( 
\begin{array}{c}
\mathbf{Q} \\ 
\mathbf{P}
\end{array}
\right) ,
\end{equation*}
and we may write the Hamiltonian as

\begin{equation*}
H=\frac{1}{2}\left( 
\begin{array}{cc}
\mathbf{Q}_{\mathbf{T}}^{T} & \mathbf{P}_{\mathbf{T}}^{T}
\end{array}
\right) \left( 
\begin{array}{cc}
\mathbf{H}_{\mathbf{Q}_{\mathbf{T}}} & \mathbf{0} \\ 
\mathbf{0} & \mathbf{H}_{\mathbf{P}_{\mathbf{T}}}
\end{array}
\right) \left( 
\begin{array}{c}
\mathbf{Q}_{\mathbf{T}} \\ 
\mathbf{P}_{\mathbf{T}}
\end{array}
\right) .
\end{equation*}

\subsubsection{The Equations of Motion:}

Since the system is decoupled for $\mathbf{Q}_{\mathbf{T}}$ and $\mathbf{P}_{%
\mathbf{T}}$, the equations of motion are easily calculated: 
\begin{equation*}
\left( 
\begin{array}{c}
\mathbf{Q}_{\mathbf{T}}(t) \\ 
\mathbf{P}_{\mathbf{T}}(t)
\end{array}
\right) =\left( 
\begin{array}{cc}
\mathbf{\Lambda }_{\mathbf{C}} & \mathbf{\Lambda }_{\mathbf{S}} \\ 
-\mathbf{\Lambda }_{\mathbf{S}} & \mathbf{\Lambda }_{\mathbf{C}}
\end{array}
\right) \left( 
\begin{array}{c}
\mathbf{Q}_{\mathbf{T}}(0) \\ 
\mathbf{P}_{\mathbf{T}}(0)
\end{array}
\right) ,
\end{equation*}
where $\mathbf{\Lambda }_{\mathbf{C}}$ and $\mathbf{\Lambda }_{\mathbf{S}}$
are $N\times N$ diagonal matrices whose elements are 
\begin{eqnarray*}
\left[ \mathbf{\Lambda }_{\mathbf{C}}\right] _{i,i} &\equiv &\cos (\Omega
_{i,i}t), \\
\left[ \mathbf{\Lambda }_{\mathbf{S}}\right] _{i,i} &\equiv &\sin (\Omega
_{i,i}t).
\end{eqnarray*}
The frequencies of the normal modes are the $\Omega _{i,i}$.

Notice that 
\begin{equation*}
\mathbf{\Lambda }\equiv \left( 
\begin{array}{cc}
\mathbf{\Lambda }_{\mathbf{C}} & \mathbf{\Lambda }_{\mathbf{S}} \\ 
-\mathbf{\Lambda }_{\mathbf{S}} & \mathbf{\Lambda }_{\mathbf{C}}
\end{array}
\right) =\mathbf{\Lambda }_{\mathbf{1}}\cdot \mathbf{\Lambda }_{\mathbf{2}%
}\cdot \ldots \cdot \mathbf{\Lambda }_{\mathbf{N-1}}\cdot \mathbf{\Lambda }_{%
\mathbf{N}},
\end{equation*}
where $\mathbf{\Lambda }_{\mathbf{i}}$ are $2N\times 2N$ matrices whose
elements are given by 
\begin{eqnarray*}
\left[ \mathbf{\Lambda }_{\mathbf{i}}\right] _{i,i} &=&\left[ \mathbf{%
\Lambda }_{\mathbf{i}}\right] _{N+i,N+i}=\cos (\Omega _{i,i}t), \\
\left[ \mathbf{\Lambda }_{\mathbf{i}}\right] _{i,N+i} &=&\sin (\Omega
_{i,i}t), \\
\left[ \mathbf{\Lambda }_{\mathbf{i}}\right] _{N+i,i} &=&-\sin (\Omega
_{i,i}t), \\
\left[ \mathbf{\Lambda }_{\mathbf{i}}\right] _{k,k} &=&1\text{ for }k\neq i%
\text{ and }k\neq N+i, \\
\left[ \mathbf{\Lambda }_{\mathbf{i}}\right] _{k,l} &=&0\text{ for the other
cases.}
\end{eqnarray*}
In this manner we may see $\mathbf{\Lambda }$ as a sequence of $N$ rotations
in the planes $q_{i_{\mathbf{T}}}\times p_{i_{\mathbf{T}}}$, where $q_{i_{%
\mathbf{T}}}$ and $p_{i_{\mathbf{T}}}$ are the elements of $\mathbf{Q}_{%
\mathbf{T}}$ and.$\mathbf{P}_{\mathbf{T}}$:

\begin{eqnarray*}
\mathbf{Q}_{\mathbf{T}}^{T} &\equiv &\left( 
\begin{array}{ccccc}
q_{1_{\mathbf{T}}} & q_{2_{\mathbf{T}}} & \cdots & q_{N-1_{\mathbf{T}}} & 
q_{N_{\mathbf{T}}}
\end{array}
\right) , \\
\mathbf{P}_{\mathbf{T}}^{T} &\equiv &\left( 
\begin{array}{ccccc}
p_{1_{\mathbf{T}}} & p_{2_{\mathbf{T}}} & \cdots & p_{N-1_{\mathbf{T}}} & 
p_{N_{\mathbf{T}}}
\end{array}
\right) .
\end{eqnarray*}

Defining 
\begin{equation*}
\mathbf{Z}\equiv \left( 
\begin{array}{cc}
\mathbf{M}_{\mathbf{T}} & \mathbf{0} \\ 
\mathbf{0} & \mathbf{M}_{\mathbf{T}}^{-1}
\end{array}
\right) \left( 
\begin{array}{cc}
\mathbf{M}_{\mathbf{R}} & \mathbf{0} \\ 
\mathbf{0} & \mathbf{M}_{\mathbf{R}}
\end{array}
\right) \left( 
\begin{array}{cc}
\mathbf{M}_{\mathbf{S}} & \mathbf{0} \\ 
\mathbf{0} & \mathbf{M}_{\mathbf{S}}^{-1}
\end{array}
\right) ,
\end{equation*}
the equations of motion for the original variables may be written as 
\begin{equation*}
\left( 
\begin{array}{c}
\mathbf{Q}\left( t\right) \\ 
\mathbf{P}\left( t\right)
\end{array}
\right) \equiv \mathbf{Z}^{-1}\cdot \mathbf{\Lambda }\cdot \mathbf{Z}\cdot
\left( 
\begin{array}{c}
\mathbf{Q}\left( 0\right) \\ 
\mathbf{P}\left( 0\right)
\end{array}
\right) .
\end{equation*}

\section{Quantum Systems}

As is well known quantum problems involving harmonic oscillators and linear
couplings are in general equivalent to the classical ones, except for
essentially quantum initial conditions. As we show here the transformations
are the same as the ones obtained in the classical case.

Consider a quantum oscillator linearly coupled to $N-1$ other oscillators.
The Hamiltonian operator in the so called \textit{rotating wave approximation%
} (RWA) shall be written as 
\begin{equation}
H=\overset{N}{\underset{i=1}{\sum }}g_{i,i}a_{i}^{\dagger }a_{i}+\overset{N}{%
\underset{i=2}{\sum }}g_{1,i}\left( a_{1}^{\dagger }a_{i}+a_{i}^{\dagger
}a_{1}\right) ,  \label{HamitQ}
\end{equation}
where $g_{i,i}>0$, $g_{i,j}\leqslant 0$ for $i\neq j$, and we set $\hbar =1$%
. The operators $a_{i}$ and $a_{i}^{\dagger }$ are bosonic creation and
annihilation operators: 
\begin{equation}
\left[ a_{i},a_{j}^{\dagger }\right] =\delta _{i,j}.  \label{comutadores}
\end{equation}

This Hamiltonian may concern a microwave mode constructed in a
superconducting cavity which is connected to $N-1$ other superconducting
cavities by waveguides, as it was assumed in Ref. \cite{art10} for $N=2$,
where no
environment was considered. Otherwise, we may regard the oscillator related
to $a_{1}$ as the system of interest and may use the other oscillators for
simulating the environment.

In the following, the matrices of Section 2 will help us to get information
about this system. The orthonormality of $\mathbf{\mathbf{M}}_{\mathbf{R}}$,
Eq. (\ref{ortonormalidade}), will be used in several steps below.

Defining 
\begin{equation*}
\mathbf{M}_{a}\mathbf{\equiv }\left( 
\begin{array}{c}
a_{1} \\ 
a_{2} \\ 
\vdots \\ 
a_{N}
\end{array}
\right) \text{ \ \ and \ \ }\mathbf{M}_{a^{\dagger }}\mathbf{\equiv }\left( 
\begin{array}{c}
a_{1}^{\dagger } \\ 
a_{2}^{\dagger } \\ 
\vdots \\ 
a_{N}^{\dagger }
\end{array}
\right) ,
\end{equation*}
the Hamiltonian (\ref{HamitQ}) may be written as 
\begin{equation*}
H=\left( \mathbf{M}_{a^{\dagger }}\right) ^{T}\cdot \mathbf{H}_{\mathbf{Q}_{%
\mathbf{S}}}\mathbf{\cdot M}_{a},
\end{equation*}
where $\mathbf{H}_{\mathbf{Q}_{\mathbf{S}}}$ is given in Eq. (\ref{HQS}).
Using Eq. (\ref{HQR}), we write the Hamiltonian in a diagonal form: 
\begin{equation*}
H=\left( \mathbf{M}_{\bar{a}^{\dagger }}\right) ^{T}\cdot \mathbf{H}_{%
\mathbf{Q}_{\mathbf{R}}}\mathbf{\cdot M}_{\bar{a}}=\overset{N}{\underset{i=1%
}{\sum }}\lambda _{i,i}\bar{a}_{i}^{\dagger }\bar{a}_{i},
\end{equation*}
where 
\begin{equation}
\mathbf{M}_{\bar{a}}\equiv \left( 
\begin{array}{c}
\bar{a}_{1} \\ 
\bar{a}_{2} \\ 
\vdots \\ 
\bar{a}_{N}
\end{array}
\right) \equiv \mathbf{M}_{\mathbf{R}}\mathbf{\cdot M}_{a}\text{ \ \ and \ \ 
}\mathbf{M}_{\bar{a}^{\dagger }}\equiv \left( 
\begin{array}{c}
\bar{a}_{1}^{\dagger } \\ 
\bar{a}_{2}^{\dagger } \\ 
\vdots \\ 
\bar{a}_{N}^{\dagger }
\end{array}
\right) \equiv \mathbf{M}_{\mathbf{R}}\mathbf{\cdot M}_{a^{\dagger }}.
\label{Ma}
\end{equation}

The operators $\bar{a}_{i}$ and $\bar{a}_{i}^{\dagger }$ above are the
bosonic operators related to the normal modes of this system, whose
frequencies are $\lambda _{i,i}$. They obey the usual commutation relations
for bosons, as may be seen by first writing the commutation relations for
the $a_{i}$ and $a_{i}^{\dagger }$ in a matricial form, 
\begin{equation*}
\mathbf{M}_{a}\mathbf{\cdot }\left( \mathbf{M}_{a^{\dagger }}\right)
^{T}-\left( \mathbf{M}_{a^{\dagger }}\mathbf{\cdot }\left( \mathbf{M}%
_{a}\right) ^{T}\right) ^{T}=\mathbf{I,}
\end{equation*}
and then observing that 
\begin{equation*}
\mathbf{M}_{\bar{a}}\mathbf{\cdot }\left( \mathbf{M}_{\bar{a}^{\dagger
}}\right) ^{T}-\left( \mathbf{M}_{\bar{a}^{\dagger }}\mathbf{\cdot }\left( 
\mathbf{M}_{\bar{a}}\right) ^{T}\right) ^{T}=\mathbf{M}_{\mathbf{R}}\mathbf{%
\cdot }\left( \mathbf{M}_{a}\mathbf{\cdot }\left( \mathbf{M}_{a^{\dagger
}}\right) ^{T}-\left( \mathbf{M}_{a^{\dagger }}\mathbf{\cdot }\left( \mathbf{%
M}_{a}\right) ^{T}\right) ^{T}\right) \mathbf{\cdot M}_{\mathbf{R}}^{T}=%
\mathbf{I.}
\end{equation*}
If $\left| 0\right\rangle $ is the vacuum state related to the original
operators $a_{i}$, we must have 
\begin{equation*}
a_{i}\left| 0\right\rangle =0\left| 0\right\rangle
\end{equation*}
for any $a_{i}$. It is easy to see that $\left| 0\right\rangle $ is the
vacuum state related to the operators $\bar{a}_{i}$, since the $\bar{a}_{i}$
are linear combinations of the $a_{i}$.

The number of excitations of the normal modes is a conserved, since 
\begin{equation*}
\left[ \overset{N}{\underset{i=1}{\sum }}\bar{a}_{i}^{\dagger }\bar{a}_{i},H%
\right] =0.
\end{equation*}
Observing that 
\begin{equation*}
\overset{N}{\underset{i=1}{\sum }}\bar{a}_{i}^{\dagger }\bar{a}_{i}=\left( 
\mathbf{M}_{\bar{a}^{\dagger }}\right) ^{T}\mathbf{\cdot M}_{\bar{a}}=\left( 
\mathbf{M}_{a^{\dagger }}\right) ^{T}\mathbf{\cdot M}_{a}=\overset{N}{%
\underset{i=1}{\sum }}a_{i}^{\dagger }a_{i},
\end{equation*}
we see that the number of excitations is conserved for the original modes
too.

Let us define\ $\left| 1_{i}\right\rangle $ as the state where there is one
excitation in the $i^{th}$ oscillator and vacuum in the other oscillators.
Thus, 
\begin{equation*}
\mathbf{M}_{\left| 1\right\rangle }\mathbf{\equiv }\left( 
\begin{array}{c}
\left| 1_{1}\right\rangle \\ 
\left| 1_{2}\right\rangle \\ 
\vdots \\ 
\left| 1_{N}\right\rangle
\end{array}
\right) \equiv \mathbf{M}_{a^{\dagger }}\left| 0\right\rangle .
\end{equation*}
Analogously, we may define one excitation states for the normal modes: 
\begin{equation*}
\mathbf{M}_{\left| \bar{1}\right\rangle }\mathbf{\equiv }\left( 
\begin{array}{c}
\left| \bar{1}_{1}\right\rangle \\ 
\left| \bar{1}_{2}\right\rangle \\ 
\vdots \\ 
\left| \bar{1}_{N}\right\rangle
\end{array}
\right) \mathbf{\equiv M}_{\bar{a}^{\dagger }}\left| 0\right\rangle .
\end{equation*}
Observing that 
\begin{eqnarray*}
e^{iHt}\left| \bar{1}_{i}\right\rangle &=&e^{i\lambda _{i,i}t}\left| \bar{1}%
_{i}\right\rangle , \\
\mathbf{M}_{\left| \bar{1}\right\rangle } &=&\mathbf{M}_{\mathbf{R}}\mathbf{%
\cdot M}_{\left| 1\right\rangle },
\end{eqnarray*}
it is easy to see that 
\begin{equation*}
e^{iHt}\mathbf{M}_{\left| 1\right\rangle }=\mathbf{M}_{\mathbf{R}}^{T}%
\mathbf{\cdot }e^{i\mathbf{H}_{\mathbf{Q}_{\mathbf{R}}}t}\mathbf{\cdot M}_{%
\mathbf{R}}\mathbf{\cdot M}_{\left| 1\right\rangle },
\end{equation*}
where 
\begin{equation*}
e^{i\mathbf{H}_{\mathbf{Q}_{\mathbf{R}}}t}=\left( 
\begin{array}{cccc}
e^{i\lambda _{1,1}t} & 0 & \cdots & 0 \\ 
0 & e^{i\lambda _{2,2}t} & \cdots & 0 \\ 
\vdots & \vdots & \ddots & \vdots \\ 
0 & 0 & \cdots & e^{i\lambda _{N,N}t}
\end{array}
\right) ,
\end{equation*}
what gives the temporal evolutions of the states $\left| 1_{i}\right\rangle $%
.

\bigskip

\section{Conclusion}

We performed a detailed analysis of\ a system composed by one oscillator
linearly coupled to $N-1$ other oscillators. An intuitive view about this
problem may be achieved: for $N=2$ and $N=3$, the variable transformations
carried out in order to find the system's normal modes may be geometrically
visualized, and for other values of $N$ analogous transformations occur. If
we consider that the central oscillator is the system of interest and the
other oscillators are the environment, we will be able to study decoherence,
 e. g.,
comparing results obtained for this system to results obtained using master
equations \cite{art7}, or even to results coming from the laboratory 
\cite{art11}.

The matricial technics developed here are easier to deal with in the
classical systems. The classical results can be almost directly used for the
quantum systems. Of course, other topologies for the oscillators may be
investigated using similar matricial technics, e. g., two oscillators
coupled to $N$ other oscillators. If we consider the two central oscillators
as the system of interest and the $N$ other as the environment, conditions
for the appearing of decoherence-free subspaces \cite{art12,art13,art14,art15}
may be
searched (two oscillators subjected to the same bath were studied in 
\cite{art8,art16, art17}
using master equations). Knowledge about this topic is important in the
quantum information context, since the main problem in this area nowadays is
the deleterious action of the environment over quantum coherences
(decoherence). Detailed comparison from our results with master equation
results will be the subject of a forthcoming publication.

\begin{acknowledgments}
The authors acknowledge financial support from the Brazilian agency
CNPq (Conselho Nacional de Desenvolvimento Cient\'{\i}fico e Tecn\'{o}logico).
\end{acknowledgments}

\appendix

\section*{Appendix}

In order to see why the decomposition of Section 2.2.1 is possible, we first
notice that it is possible for $N=2$: 
\begin{equation*}
\mathbf{M}_{\mathbf{R}}=\mathbf{R}_{\mathbf{1}}=\mathbf{R}_{1,2}=\left( 
\begin{array}{cc}
\cos \alpha _{1,2} & \sin \alpha _{1,2} \\ 
-\sin \alpha _{1,2} & \cos \alpha _{1,2}
\end{array}
\right) \text{ \ \ \ for \ \ \ }N=2.
\end{equation*}
Then we prove the general case showing that if the decomposition is possible
for a $\left( N-1\right) $ $\times \left( N-1\right) $ orthogonal matrix it
will be possible for a $N\times N$ orthogonal matrix.

Consider a $N\times N$ real matrix $\mathbf{A}$. Observing that for $k\neq 1$
and $k\neq j$%
\begin{equation*}
\left[ \mathbf{A\cdot R}_{1,j}^{T}\right] _{1,k}=\left[ \mathbf{A}\right]
_{1,k},
\end{equation*}
\bigskip and that it is always possible to find a value for $\alpha _{1,j}$
which yields 
\begin{equation*}
\left[ \mathbf{A\cdot R}_{1,j}^{T}\right] _{1,j}=0,
\end{equation*}
we see that it is possible to chose all the $\alpha _{1,j}$ in such a way
that 
\begin{eqnarray*}
\mathbf{B} &\equiv &\mathbf{M}_{\mathbf{R}}\cdot \mathbf{R}_{1,2}^{T}\cdot
...\cdot \mathbf{R}_{1,N-1}^{T}\cdot \mathbf{R}_{1,N}^{T} \\
&\equiv &\left( 
\begin{array}{cccc}
1 & 0 & \cdots  & 0 \\ 
B_{2,1} & B_{2,2} & \cdots  & B_{2,N} \\ 
\vdots  & \vdots  & \ddots  & \vdots  \\ 
B_{N,1} & B_{N,2} & \cdots  & B_{N,N}
\end{array}
\right) .
\end{eqnarray*}
We may write 
\begin{equation}
\mathbf{M}_{\mathbf{R}}=\mathbf{B}\cdot \mathbf{\mathbf{R}_{\mathbf{1}}.}
\label{MR=B.R1}
\end{equation}

Observing that if \bigskip $\mathbf{A}$ is orthogonal then $\mathbf{A}\cdot 
\mathbf{R}_{1,j}^{T}$ is orthogonal too, we see that $\mathbf{B}$ has the
form 
\begin{equation*}
\mathbf{B}\equiv \left( 
\begin{array}{cc}
1 & 0\cdots 0 \\ 
\begin{array}{c}
0 \\ 
\vdots \\ 
0
\end{array}
& \mathbf{C}
\end{array}
\right) ,
\end{equation*}
where $\mathbf{C}$ is a $\left( N-1\right) \times \left( N-1\right) $
orthogonal matrix. Since we are assuming that the decomposition proposed is
valid for a $\left( N-1\right) \times \left( N-1\right) $ orthogonal matrix,
we may write 
\begin{equation*}
\mathbf{C=D}_{\mathbf{N-2}}\cdot \mathbf{D}_{\mathbf{N-3}}\cdot ...\cdot 
\mathbf{D}_{\mathbf{2}}\cdot \mathbf{D}_{\mathbf{1}},
\end{equation*}
where the $\mathbf{D}_{\mathbf{i}}$ are the $\left( N-1\right) \times \left(
N-1\right) $ matrices analog to the $\mathbf{R}_{\mathbf{i}}$. Thus 
\begin{equation*}
\mathbf{B}\equiv \left( 
\begin{array}{cc}
1 & 0\cdots 0 \\ 
\begin{array}{c}
0 \\ 
\vdots \\ 
0
\end{array}
& \mathbf{D}_{\mathbf{N-2}}
\end{array}
\right) \cdot \left( 
\begin{array}{cc}
1 & 0\cdots 0 \\ 
\begin{array}{c}
0 \\ 
\vdots \\ 
0
\end{array}
& \mathbf{D}_{\mathbf{N-3}}
\end{array}
\right) \cdot ...\cdot \left( 
\begin{array}{cc}
1 & 0\cdots 0 \\ 
\begin{array}{c}
0 \\ 
\vdots \\ 
0
\end{array}
& \mathbf{D}_{\mathbf{2}}
\end{array}
\right) \cdot \left( 
\begin{array}{cc}
1 & 0\cdots 0 \\ 
\begin{array}{c}
0 \\ 
\vdots \\ 
0
\end{array}
& \mathbf{D}_{\mathbf{1}}
\end{array}
\right) .
\end{equation*}

Notice that we may regard 
\begin{equation*}
\mathbf{R}_{\mathbf{i+1}}=\left( 
\begin{array}{cc}
1 & 0\cdots 0 \\ 
\begin{array}{c}
0 \\ 
\vdots \\ 
0
\end{array}
& \mathbf{D}_{\mathbf{i}}
\end{array}
\right) ,
\end{equation*}
and then 
\begin{equation*}
\mathbf{B}\equiv \mathbf{R}_{\mathbf{N-1}}\cdot \mathbf{R}_{\mathbf{N-2}%
}\cdot ...\cdot \mathbf{R}_{\mathbf{3}}\cdot \mathbf{R}_{\mathbf{2}}.
\end{equation*}
Using equation (\ref{MR=B.R1}) we have $\mathbf{M}_{\mathbf{R}}$ in the form
given by (\ref{MR=RN-1...R1}).

\newpage

\newpage

\newpage

\newpage

\newpage

\newpage

\end{document}